\pdfoutput=1

\documentclass[11pt]{article}

\usepackage[final]{acl}

\usepackage{times}
\usepackage{latexsym}

\usepackage[T1]{fontenc}

\usepackage[utf8]{inputenc}

\usepackage{microtype}

\usepackage{amsmath}
\usepackage{amssymb}
\usepackage{mathtools}
\usepackage{amsthm}
\usepackage{amsfonts}
\usepackage[capitalize,noabbrev]{cleveref}
\usepackage{listings}

\usepackage{inconsolata}

\usepackage{graphicx}
\usepackage{listings}
\usepackage{booktabs} 
\title{CodeSSM: Towards State Space Models for Code Understanding}

\author{
 \textbf{Shweta Verma\textsuperscript{1, *}} ,
 \textbf{Abhinav Anand\textsuperscript{1, *}} ,
 \textbf{Mira Mezini\textsuperscript{1,2,3}}
 \\
 \textsuperscript{1}TU Darmstadt,
 \textsuperscript{2}Hessian Center for Artificial Intelligence, Darmstadt, Germany,
 \\
 \textsuperscript{3}National Research Center for Applied Cybersecurity ATHENE
 \\
  \textsuperscript{*}Equal Contribution,
\\
 \small{
   \textbf{Correspondence:} \href{mailto:email@domain}{shweta.verma@tu-darmstadt.de}
 }
}

\begin{document}

\maketitle
\begin{abstract}


Although transformers dominate many code-specific tasks, they have significant limitations. This paper explores State Space Models (SSMs) as a promising alternative for code understanding tasks such as retrieval, classification, and clone detection. We introduce CodeSSM, the first SSM-based model trained on code corpora to assess its effectiveness. Our results demonstrate that SSMs are more sample-efficient and can extrapolate to longer contexts beyond the pretraining length. Extensive experiments show that SSMs offer a viable alternative to transformers, addressing several their limitations. Additionally, CodeSSM reduces memory usage by up to 64\% compared to transformers at a context length of 2048, with greater savings as context length grows.


\end{abstract}

\section{Introduction}

Transformers \citep{attentionisallyouneed} have been known to perform well in various applications. 
 One of the main reasons for this improvement is the pretraining - fine-tuning paradigm used to train transformer models, which is possible due to the ease of parallelization of attention-based transformer architectures. Under this paradigm, the transformer model is initially trained on a large corpus of unlabeled data using a self-supervised training objective, followed by a supervised fine-tuning process on a smaller labeled dataset. 

However, the performance gain offered by the transformer models comes with some trade-offs, such as quadratic complexity, a substantial data requirement, and high inference costs. Although methods such as linear attention \citep{linear-attention} and sparse attention \citep{longt5, bigbird, lsg} have been proposed to address computational inefficiency, the practical gains from these methods are limited \citep{limitation-efficient-transformer, devil-in-linear-transformer}. 

Another limitation of transformers is their fixed context window resulting from positional embedding. Although methods such as AliBi \citep{alibi} and RoPE \citep{Rope} solve the problem of a fixed context window, transformer models still fail to generalize to lengths not seen during pretraining \citep{yarn, positionalbias}. To overcome this challenge, additional techniques are required to extend the context window without significantly compromising performance \citep{lim-rope, lim-rope-fourier, lim-rope-mix}.  

One-dimensional Convolutional Neural Networks (1D CNNs) are a fast and positional embedding-free alternative to transformers. 1D CNNs, when trained using the pretraining-finetuning approach, show improved performance on some tasks, but do not capture long-range dependencies in input data \citep{pre-trained-conv}. 

\begin{figure}[!t]
\centerline{\includegraphics[width=8cm]{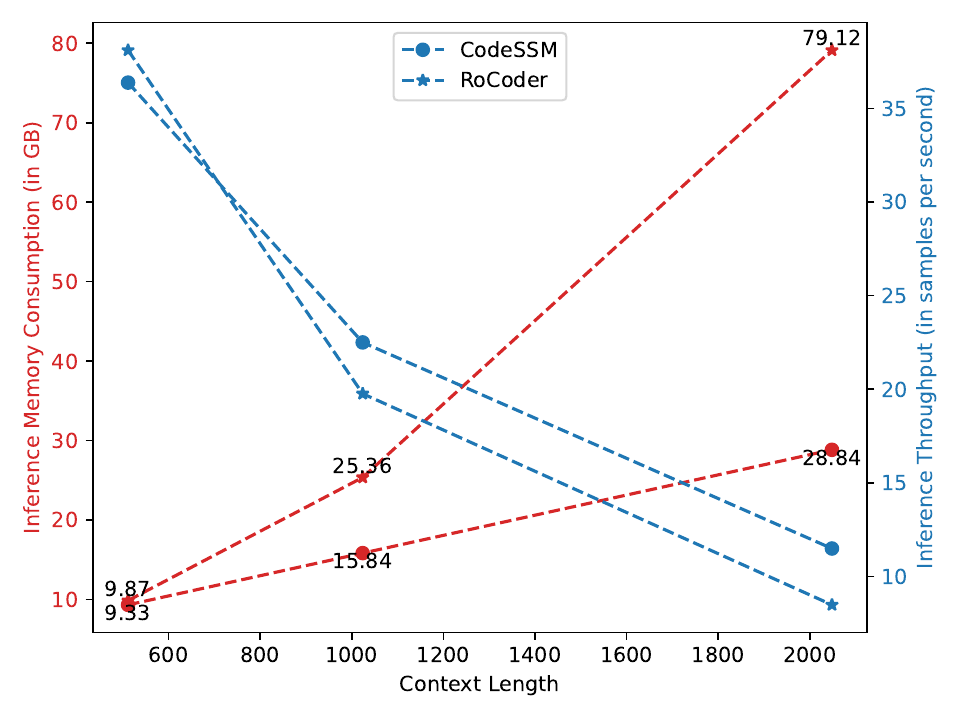}}
\caption{Memory consumption and throughput of CodeSSM and RoCoder model during inference. }
\label{mem_thr}
\end{figure}
State-space models (SSMs)\cite{linear-ssm, s4, s4d} represent an alternative that addresses the limitations of transformers and 1D CNNs. SSMs leverage the strengths of 1D CNNs while effectively capturing long-range dependencies with linear-time complexity. In addition, SSMs are position-aware, eliminating the need for positional embedding. 


In this work, we explore the application of SSMs to the domain of code understanding. To achieve this, we pretrain an encoder-only model based on SSM, which we call CodeSSM, on a small code dataset with masked language modeling (MLM) \citep{bert}. We evaluated our model on multiple code understanding benchmarks. We also investigate various training aspects, such as the effects of positional embedding, dropout, bidirectional SSMs, and varying pretraining context lengths. We compare CodeSSM with two transformer models that we train -- (1) BertCoder: based on the Bert architecture \citep{bert} and (2) RoCoder: based on the RoFormer architecture \cite{Rope}. Moreover, we compare our model with Zamba, a hybrid SSM transformer model based on the Zamba2 architecture \cite{zamba2}. We also assess the performance of our model against several high-performing transformers, which are trained with larger data and complex pretraining objectives. 

Our study demonstrates that SSMs exhibit strong sample efficiency in code
modeling tasks. In particular, CodeSSM achieves more than 50\% precision in Masked Language Modeling (MLM) with fewer than 3,000 training steps. Furthermore, we show that CodeSSM consumes up to 64\% less memory compared to transformers and is significantly faster (see \cref{mem_thr}). The efficiency in terms of speed and memory consumption in CodeSSM (compared to transformers) improves with increasing input length. 

Additionally, the code language modeling capabilities of CodeSSM also transfer well to downstream tasks, as CodeSSM outperforms transformers on multiple code understanding tasks when pretrained under similar conditions. CodeSSM is also competitive with (and even surpasses) transformer models trained on significantly larger datasets and with more complex pretraining objectives. Furthermore, another advantage of SSMs is their ability to extrapolate to much larger contexts than those encountered during pretraining, effectively addressing a key limitation of the transformer architecture.

Our goal in training the CodeSSM model is not to achieve state-of-the-art results on coding benchmarks, but to explore and document the strengths and limitations of SSMs for code understanding. This motivated our use of a small dataset and MLM pretraining, enabling rapid experimentation with and comparison of different model configurations under consistent training conditions. 
Although we observed that larger pretraining datasets improved performance, the investigation of extensive datasets and more complex training objectives is left for future work.

\noindent
\textbf{Summary of contributions:} 
\begin{itemize}
    \item 
    We present the first systematic evaluation of State-Space Models (SSMs) for code understanding, introducing a specialized model, which we term CodeSSM. Our analysis reveals that CodeSSM achieves robust performance on retrieval, classification and clone detection tasks, consistently outperforming attention-based models while requiring substantially less pretraining data and memory.
    
    \item 
    Our study presents the first empirical evidence that SSMs achieve superior sample efficiency in code-understanding tasks compared to transformers. CodeSSM substantially outperforms transformer baselines in data-constrained scenarios with only a few thousand training samples.
    
    \item 
    We demonstrate that CodeSSM addresses a key limitation of transformer architectures -- extrapolation to unseen context lengths. By examining the effects of different pretraining context lengths and positional embeddings in CodeSSM, we show that SSMs outperform transformers as downstream input sequences become longer.

    \item We observe that including even a small dropout mask in the CodeSSM layer degrades the performance of the model even on very small benchmarks. CodeSSM also performs very well on the adversarial text-code search task.
    Together, these observations indicate that SSMs exhibit robust semantic understanding capabilities.
\end{itemize}
The code is available \href{https://github.com/abx04/CodeSSM}{here}.

\section{CodeSSM}
\begin{figure}[ht]
\vskip 0.2in
\begin{center}
\centerline{\includegraphics[width=\columnwidth, page=1, viewport=0.49 8 210 319, clip]{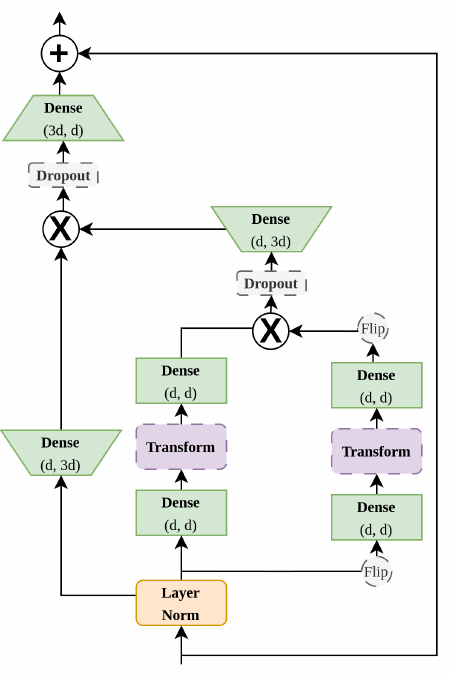}}
\caption{CodeSSM layer. Transform represents SSMs in CodeSSMs and Discrete Fourier Transforms in CodeF. Dropout is only present in one of the CodeSSM variants, CodeSSM-do. In CodeSSM-uni, both the flip operations are removed. Non-linearity is not shown for clarity.}
\label{bigs_arch}
\end{center}
\vskip -0.2in
\end{figure}
In this section, we introduce the basic architecture of CodeSSM and its variations that we investigate. We also explain the pretraining setup.
\subsection{Architecture} \label{variants}
CodeSSM is an encoder-only model consisting of 12 layers\footnote{The trained weights are available here: \url{https://figshare.com/s/14238287e9078f92cd50}.}. This model is built upon the Bidirectional Gated SSM (BiGS) architecture \cite{bigs}, incorporating certain variations that we examine and their applicability in the field of code understanding.

\textbf{CodeSSM-pos.} Transformer models \cite{attentionisallyouneed} require positional embeddings to address their permutation invariance in self-attention layers. 
Since SSMs are inherently position-aware, they do not require positional embeddings. However, to understand the implications of positional embeddings on downstream performance, we include them in one variant of CodeSSM, which we call CodeSSM-pos. 

\textbf{CodeSSM-dropout.}
Dropout layers \citep{dropout} prevent overfitting in neural networks. So, to investigate overfitting in CodeSSM, we create a variant with two dropout masks, positioned after each multiplication operation (see \cref{bigs_arch}). We refer to this variant as CodeSSM-do.

\textbf{CodeF and CodeSSM-Uni.}
Recent research questioned the necessity of self-attention in transformers \cite{fnet}. 
To investigate the significance of SSM, we conducted a similar ablation study by replacing SSM blocks in CodeSSM with DFT (see \cref{bigs_arch}), creating a variant we call CodeF.
Furthermore, we analyze the importance of bidirectional processing via CodeSSM-Uni (Unidirectional Gated SSM), a variant of CodeSSM that removes flip operations from the CodeSSM layer.

\subsection{Pretraining} \label{pretraining}
We trained all models with the MLM objective using the MLM setup proposed by \citet{bert}.
We use different pretraining setups to investigate various aspects of SSMs on code understanding tasks. Here, we explain this setup. 

\textbf{Dataset.} We initially trained CodeSSM and a BERT-like transformer model, which we refer to as BertCoder, on 1.8M text code pairs from the CodeSearchNet (CSN) dataset. The small pretraining dataset allows us to quickly experiment with various model configurations. Both models were trained with a context window of 2048, enabling them to be fine-tuned on benchmarks with lengths shorter than 2048. We refer to the CodeSSM model trained with the CSN dataset as CodeSSM-base. 
We also train CodeSSM variants such as CodeSSM-pos, CodeSSM-do, CodeF and CodeSSM-Uni (see \ref{variants}) on the same dataset and context length.

\textbf{Context Length and Length Extrapolation} To investigate the length extrapolation capability of SSMs, we pre-train CodeSSM with a context window of 256 length. Since we cannot change the context window for BertCoder once pretrained with a smaller context, we pretrained a RoFormer model \cite{Rope}, which is similar to the BERT model but with Rotatory Positional Embeddings (RoPE), with a context window of length 256. We refer to this model as RoCoder.

The CSN dataset separates the docstring and code, which creates an issue when they are concatenated and subsequently truncated. This can result in one modality (text) being much longer than the other modality (code). We also observed that this discrepancy results in a reduction in performance in tasks involving two sequences (refer to column 3 of \cref{codexglue-table}). 

To alleviate this issue, we use a subset of 1.8M samples from the StarCoder dataset \citep{starcoder} to train models with 256-length context (referred to as CodeSSM-sc-256). This subset contains 300k samples from the six programming languages (PLs) in the CSN dataset \footnote{Go, Java, Javascript, PHP, Python, Ruby}. We also train CodeSSM on this dataset with a context length of 2048 (referred to as CodeSSM-sc-2048).

\textbf{Data Scaling.} To study the impact of large datasets on the downstream performance of CodeSSM, we also train it in a large subset of the StarCoder dataset with 4.8M samples (referred to as CodeSSM-sc-L-2048). This subset contains 600k samples from 8 PLs (C and C\# along with CSN PLs).

\begin{table}[t]
\caption{Comparison of dataset size and pretraining objective of various encoder-only models. Replaced Token Detection (RTD), Unidirectional Language Modeling (ULM), Denoising objective (DNS), Data Flow Edge Prediction (DFEP), Node Alignment (NA), Masked Span Prediction (MSP), Identifier Tagging (IP), Masked Identifier Prediction (MIP) }
\label{modelanddata}
\vskip 0.15in
\begin{center}
\begin{tiny}
\begin{sc}
\begin{tabular}{lcccr}
\toprule
Models & Dataset  & Pretraining \\
 &  Size & Objectives\\
\midrule
CodeSSM-base  & 1.8M CSN & MLM \\
CodeSSM-SC  & 1.8M SC data & MLM \\
CodeSSM-SC-L  & 4.8M SC data & MLM \\
CodeBERT   & 8.5M CSN & MLM, RTD  \\
UniXCoder & 6.4M CSN + C4 data  & MLM, ULM, DNS  \\
GraphCodeBERT  & 6.4M CSN & MLM, DFEP \\
CodeT5  & 8.35M CSN + C, CSharp &  MSP, IP, MIP\\

\bottomrule
\end{tabular}
\end{sc}
\end{tiny}
\end{center}
\vskip -0.1in
\end{table}
\section{Experiments}
\label{sec: experiments}
In this section, we describe baselines to which we compare CodeSSM and the benchmarks on which we evaluate it.

\subsection{Baselines}
We trained two transformer-based models as baselines: a large context (2048) BERT model to compare SSMs' large context abilities and a small context RoFormer model to compare the length extrapolation abilities of transformers and SSMs. We also train a Zamba model to compare our model with the other SSM variants. The models are trained in the same settings: the same data set, the same objective, the same context length, and batch size.

Additionally, we also mention the performance of some well-performing transformer models in each task to place CodeSSM in the context of SOTA encoder-only transformer models. However, it should be noted that these transformer models are trained on a much larger dataset and with a mixture of multiple training objectives (see \cref{modelanddata}).

\subsection{Benchmarks}\label{bench}
We benchmark the models on seven code understanding tasks: two retrieval tasks, three sequence classification tasks, one token-level classification task, and one clone detection task.

\textbf{Retrieval tasks.}
There are two retrieval tasks on which we evaluate our models, NLCodeSearch (AdvTest) \cite{codexglue} and Stackoverflow Question Answer (SQA).
In NLCodeSearch, a description in natural language is provided, and the goal is to find the source code that corresponds to that description. To evaluate a model's ability to generalize, function names and variable names in the test sets are replaced with generic tokens, for example, the function name with \texttt{func}. This task is difficult because the model has to locate the correct code throughout the test set. In SQA, a StackOverflow question is given, and the goal is to retrieve the highest upvoted answer. This dataset is derived from the original StackOverflow dataset \footnote{\url{https://www.kaggle.com/datasets/stackoverflow/stacksample/data}}. The average lengths of question and answer are 1.2k and 1.4k, respectively. 

\textbf{Sequence classification tasks.}
Sequence classification includes two vulnerability detection tasks, Devign \cite{devign} and DiverseVul \cite{diversevul}, as well as a complexity prediction task \cite{complexity}. Vulnerability detection is a binary classification task in which the goal is to identify whether a given code contains security vulnerabilities. The Devign benchmark \cite{devign} is a balanced dataset, while DiverseVul consists of only around 5\% vulnerable samples. So, we use accuracy for the former and F1 macro for the latter. Complexity prediction \cite{complexity} is a small benchmark with only 3613 training samples to predict the algorithmic complexity of a given Java code among 7 different labels. 

\textbf{Token classification task.}
In the type-inference task \cite{Type-inference}, the model has to predict the type (such as integer, boolean, etc.) of each token in a given code. The label consists of 50K most frequent (built-in and user-defined) types, while the remaining types have been replaced with the \texttt{UNK} type. During the evaluation, predicting the \texttt{UNK} token is considered incorrect.

\textbf{Clone detection task.}
In the clone detection task \cite{clone}, two pieces of code are given as input, and the task is to perform binary classification to determine whether they are semantically equivalent. This task is challenging due to the highly unbalanced dataset, which contains a significantly high number of semantically dissimilar samples. The models are evaluated using the F1 score. 

For all tasks, we maintain the same context length as specified in the original benchmarks. All models are fine-tuned using a context length of 512 for text-to-code search, clone detection, type inference, and DiverseVul. For the Devign benchmark, the models are fine-tuned with a context length of 400. In SQA and complexity prediction, we use a context length of 1024.

\begin{table*}[t] \centering
\caption{Results on NLCodeSearch (AdvTest), Clone detection and Type inference. The result also includes details about the pretraining dataset and pretraining context length. The best performance is in bold and other noteworthy results are underlined.}
\label{codexglue-table}
\vskip 0.15in
\begin{center}
\begin{tiny}
\begin{sc}
\begin{tabular}{lccccccr}
\toprule
Dataset & Model & Context Length & NLCodeSearch  & Clone Detection & Type Inference\\
 (Pretraining)  & & (Pretraining) & (MRR) & (F1) & (Overall F1)\\
\midrule
&CodeBERT   & 512 & 27.19 &  0.941 & 0.595\\
&BertCoder & 2048  & 24.30 &  \underline{0.942} & \underline{0.666} \\
CodeSeachNet & CodeSSM-base  & 256 & 21.87  & 0.908 & 0.588\\
&CodeSSM-base  & 2048 & \underline{28.38} & 0.935 & 0.615\\
&CodeSSM-pos & 2048 & 26.58  & 0.939 & 0.594\\
&CodeSSM-do & 2048 & 26.33 &  0.931 & 0.609\\
\hline
& RoCoder & 256  & 29.44 & 0.938 & \textbf{0.687} \\
& Zamba & 256  & - & 0.232 & 0.503 \\
StarCoder & CodeSSM-Sc  & 256 & 28.41 & \textbf{0.948} & 0.624\\
& CodeSSM-Sc  & 2048 & \underline{29.88} &  0.938 & 0.625\\
& CodeSSM-Sc-L  & 2048 & \textbf{30.87} & \underline{0.940} & \underline{0.637}\\
\bottomrule
\end{tabular}
\end{sc}
\end{tiny}
\end{center}
\vskip -0.1in
\end{table*}

\begin{table}[t] \centering
\caption{Results on SQA benchmark in terms of MRR. The best performance is in bold and other noteworthy results are underlined.}
\label{sqa-apps-table}
\vskip 0.15in
\begin{center}
\begin{tiny}
\begin{sc}
\begin{tabular}{lccccccr}
\toprule
Model & SQA  \\
\midrule
BertCoder & 76.25  \\
RoCoder & 80.91  \\
Zamba &  - \\
CodeSSM-Sc-256  & \underline{82.12}  \\
CodeSSM-Sc-2048  & 82.29 \\
CodeSSM-Sc-L-2048  & \textbf{82.59}\\
\bottomrule
\end{tabular}
\end{sc}
\end{tiny}
\end{center}
\vskip -0.1in
\end{table}

\section{Results} \label{performance}

We present the performance of our model on various benchmarks described in \cref{bench}

\subsection{Retrieval Tasks}

\textbf{NLCodeSearch(AdvTest).}
Column 3 of \cref{codexglue-table} presents the results of various models evaluated in NLCodeSearch benchmark. Among the models trained on the CSN dataset, CodeSSM-base (with a context size of 2048) outperforms other models, including CodeBERT, which was pretrained with significantly more data. CodeSSM-base with a context length of 256 performs poorly compared to other variants -- we explain the reason for this in \cref{pretraining}.
CodeSSM-sc-256 shows remarkable performance and length extrapolation to double the length on which it was trained and is competitive with RoCoder and CodeSSM-sc-2048. However, Zamba -- the other SSM based model -- failed to train on retrieval task. We attribute the inability to train on retrieval-based tasks to the causal convolution in the Mamba block, which prevents bidirectional understanding.

These results are particularly noteworthy given the adversarial nature of the NLCodeSearch benchmark, where critical semantic identifiers such as function and variable names are obfuscated with generic placeholder tokens. The superior performance of CodeSSM under these challenging conditions underscores its efficiency in capturing the underlying structure of code, independent of specific identifier names, even if it is pretrained with much less data compared to CodeBERT.

\textbf{StackOverflowQA.} \cref{sqa-apps-table} presents the performance of CodeSSM in conjunction with the baseline models trained by us on the SQA benchmark -- BertCoder and RoCoder. SQA allows us to evaluate the models with longer contexts. CodeSSM-sc with a pretraining context length of 256 achieves strong results: it successfully extrapolates to contexts four times longer than those seen during pretraining. Furthermore, CodeSSM-sc outperforms BertCoder (pretrained with a context length of 2048) by 6 points, despite being pretrained with a context length eight times shorter. CodeSSM also surpasses RoCoder, highlighting its superior ability to extrapolate to longer contexts and capture long-range dependencies. Furthermore, we observe that increasing the pretraining context length leads to further improvements in CodeSSM's performance. Similarly to NLCodeSearch, Zamba did not train on SQA as well. 

\begin{table}[t] \centering
\caption{Results on DiverseVul benchmark in terms of F1 macro. The best performance is in bold and other noteworthy results are underlined.}
\label{div-table}
\vskip 0.15in
\begin{center}
\begin{tiny}
\begin{sc}
\begin{tabular}{lccccccr}
\toprule
Model & DiverseVul & Devign \\
 & (F1 macro) & (Accuracy) \\
\midrule
CodeBERT   & 37.85 & 62.08\\
GraphCodeBERT   & 36.79 & 63.21 \\
CodeT5 & 45.69 & 62.88 \\
BertCoder &  65.97 & 62.50\\
RoCoder & 68.15 & \underline{53.47}\\
Zamba & 59.43 & 56.55 \\
CodeSSM-Sc-256  & \textbf{69.31} & \textbf{64.53}\\
CodeSSM-Sc-2048  & 67.72 & 63.14\\
CodeSSM-Sc-L-2048  & \underline{68.76} & 63.54\\
\bottomrule
\end{tabular}
\end{sc}
\end{tiny}
\end{center}
\vskip -0.1in
\end{table}

\begin{table}[ht] \centering
\caption{Results on Complexity Prediction. The best performance is in bold and other noteworthy results are underlined.}
\label{complexity-table}
\vskip 0.15in
\begin{center}
\begin{tiny}
\begin{sc}
\begin{tabular}{lccccccr}
\toprule
Model & F1 \\
\midrule
CodeBERT   & 85.81 \\
GraphCodeBERT   & 87.98 \\
UnixCoder   & \underline{93.75} \\
BertCoder &  89.66 \\
RoCoder &  93.06 \\
Zamba &  90.16 \\
CodeSSM-do-2048 & 92.82\\
CodeSSM-Sc-256  & \underline{93.77}\\
CodeSSM-Sc-2048  & 94.37\\
CodeSSM-Sc-L-2048  & \textbf{94.55}\\
\bottomrule
\end{tabular}
\end{sc}
\end{tiny}
\end{center}
\vskip -0.1in
\end{table}

\subsection{Sequence classification tasks}

\textbf{Devign.} Column 3 of \cref{div-table} shows the results of the various models on the devign benchmark. CodeSSM outperforms BertCoder and RoCoder as well as CodeBERT, GraphCodeBERT, and CodeT5.  
Interestingly, BertCoder, pretrained with a longer context, also outperforms CodeBERT, showcasing the importance of a longer context for transformers. However, in the case of CodeSSM, the smaller context model performs much better. Notably, RoCoder performs poorly, potentially due to overfitting, as the devign dataset is very small (only 21k training samples). RoPE, used in RoCoder, is also prone to positional overfitting \cite{lost-in-the-middle, positionalbias}. Despite being a unidirectional model, Zamba performs slightly better than RoCoder.

\textbf{DiverseVul.} Column 2 of \cref{div-table} shows that CodeSSM and its variants outperform BertCoder and RoCoder on the DiverseVul benchmark, which is a larger vulnerability detection dataset compared to Devign. It shows that CodeSSM can perform better than transformers even on a larger dataset. CodeSSM even outperforms CodeBERT, GraphCodeBERT, and CodeT5, which are trained on a much larger dataset and have multiple pretraining objectives (see \cref{modelanddata}).

\textbf{Complexity Prediction.} \cref{complexity-table} shows that CodeSSM-sc-256 outperforms all transformer models, even those trained with a longer context length (BertCoder) and larger data and more pretraining objectives (CodeBERT, GraphCodeBERT, UnixCoder). Additionally, CodeSSM-sc-2048 shows very strong performance as pretraining on longer context length improves the performance further on this task. 

\subsection{Token classification task (Type inference)}
Column 5 of \cref{codexglue-table} shows the results on the type inference benchmark. 
CodeSSM (across all context lengths) matches or exceeds CodeBERT. 
However, both RoCoder and BertCoder outperform CodeSSM. 
We provide reasons for the poor performance of CodeSSM in type inference in \cref{SSM_fail} and \cref{failurecase}.

\subsection{Clone Detection}
Column 4 of \cref{codexglue-table} shows the F1 in the clone detection task. The CodeSSM-sc with a pretraining context length of 256 outperforms both CodeBERT, BertCoder, and RoCoder in terms of F1.  Under a similar training setup (2048 context in CSN), removing positional embedding does have a minimal impact on performance, but with the advantage that the maximum input length need not be fixed during pretraining. The decrease in performance after introducing dropout shows that the model does not overfit despite the data set being unbalanced.

\section{Research Questions}
In this section, we discuss some important research questions related to CodeSSM and its characteristics.

\subsection{How sample efficient is CodeSSM?} \label{modeling}

\begin{figure}[ht]
\vskip 0.2in
\begin{center}
\centerline{\includegraphics[width=\columnwidth, page=1, viewport=0.003 0.003 .71 .36, clip]{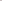}}
\caption{MLM training accuracy of CodeSSM and BertCoder, trained with a context window of 2048. The CodeSSM model achieves decent MLM accuracy in very few training steps.}
\label{bigs_bert_rope}
\end{center}
\vskip -0.2in
\end{figure}

\cref{bigs_bert_rope} shows the MLM training accuracy trajectories of CodeSSM and BertCoder, both trained with a maximum context of 2048.
CodeSSM demonstrates rapid initial learning, achieving more than 50\% accuracy in only 3,000 training steps (96,000 training samples). BertCoder, on the other hand, exhibits better long-term learning characteristics, outperforming CodeSSM by 4\% points after six training epochs (see \cref{pretrain_acc}). 
However, CodeSSM outperforms transformers in all downstream tasks except type inference, as discussed in \cref{performance}. Performance in downstream tasks shows that the code modeling capability observed during pretraining transfers well to multiple downstream tasks.

We evaluated CodeSSM on two small benchmarks: The Devign vulnerability detection benchmark with only 21k training samples and the complexity prediction benchmark with only 3.6k training samples. In both benchmarks, CodeSSM not only outperforms BertCoder and RoCoder but also many other models trained with significantly more data and multiple code-based training objectives. 

Our analyses demonstrate that SSMs represent a \textbf{sample-efficient alternative to transformers} for code language modeling and understanding tasks. 
We attribute the sample efficiency of CodeSSM to the distinct inductive bias of SSM models compared to transformer architectures \cite{bigs}. 
We also present some (informal) theoretical basis for CodeSSM's performance and the magnitude and phase spectrum of Fourier transform of SSM kernel in \cref{theory}. 

\subsection{How well does CodeSSM extrapolate to longer contexts?}
We studied transformer models and CodeSSMs for positional embedding and different pretraining context lengths to investigate their length extrapolation capabilities. 

On multiple tasks, we observe that BertCoder outperforms CodeBERT despite less pretraining data simply due to the longer pretraining context length. 
However, replacing absolute positional embedding with RoPE can improve performance even with a pretraining context of only 256. But even with RoPE, RoCoder fails to match the performance of CodeSSM except in type inference and NLCodeSearch. 

Out of two retrieval tasks, RoCoder performs slightly better on NLCodeSearch while CodeSSM outperforms it on SQA. The context length is 512 for the former and 1024 for the latter. Thus, CodeSSM is better at handling unseen input lengths as the context for the downstream task increases. Since SSMs have linear complexity with respect to input length, they are advantageous in terms of both computational efficiency and performance. Moreover, with a longer pretraining context, CodeSSM outperforms RoCoder on NLCodeSearch as well. Therefore, with more computational resources, CodeSSM can be efficiently trained with longer contexts to improve the performance on retrieval tasks. However, it is important to note that longer pretraining contexts do not benefit all downstream tasks; in fact, performance slightly decreases in vulnerability detection and clone detection.

Furthermore, we observe that \textbf{CodeSSM performs better without positional embedding}. CodeSSMs without positional embeddings do not face any length limitation, unlike transformers with absolute positional embedding, or positional biases observed with different methods such as Alibi or RoPE \citep{lost-in-the-middle, positionalbias}.


\subsection{How fast and memory efficient is CodeSSM?}
\begin{table}[ht] \centering
\caption{Memory Consumption during Inference (in GB) on various tasks with varying context lengths.}
\label{memory-table}
\vskip 0.15in
\begin{center}
\begin{tiny}
\begin{sc}
\begin{tabular}{lccccccr}
\toprule
Model &  NLCodeSearch (512) & SQA (1024) & SQA (2048)\\
\midrule
CodeSSM & 9.33 & 15.84 & 28.84 \\
RoCoder & 9.87 & 25.36 & 79.12  \\

\bottomrule
\end{tabular}
\end{sc}
\end{tiny}
\end{center}
\vskip -0.1in
\end{table}

The usage of GPU memory during inference is shown in \cref{memory-table}. CodeSSM model consumes 5.5\% less memory than the transformer model (RoCoder) on the NLCodeSearch task where the context length is 512. The difference in memory consumption between CodeSSM and RoCoder increases to 37.5\% on SQA task where the context length is 1024. As the input length increases to 2048, CodeSSM saves 63.55\% memory compared to the transformers. These findings are in line with previous works on SSM \cite{locost}.

\begin{table}[ht] \centering
\caption{Inference throughput (samples per second) with varying context lengths.}
\label{throughput-table}
\vskip 0.15in
\begin{center}
\begin{tiny}
\begin{sc}
\begin{tabular}{lccccccr}
\toprule
Model &  NLCodeSearch (512) & SQA (1024) & SQA (2048)\\
\midrule
CodeSSM & 36.38 & 22.50 & 11.49 \\
RoCoder & 38.11 & 19.76 & 8.48  \\

\bottomrule
\end{tabular}
\end{sc}
\end{tiny}
\end{center}
\vskip -0.1in
\end{table}
Transformers are slightly faster compared to SSM in small input sizes (<=512) but SSMs are faster as the input length increases. 

\subsection{Can SSMs be replaced by DFT?}
We ask this question because \citet{fnet} showed that self-attention can be replaced with DFT in transformers with minimal reduction in NLP tasks. We also observed that in all tasks RoCoder performed significantly better than BertCoder. \citet{FoPE} showed that RoPE also performs token-level (nonuniform) DFT. Additionally, SSMs also implicitly perform DFT of input and kernel during convolution operation.

\begin{figure}[ht]
\vskip 0.2in
\begin{center}
\centerline{\includegraphics[width=\columnwidth, page=1, viewport=0.003 0.003 .71 .36, clip]{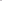}}
\caption{MLM training accuracy for different variations to the SSM layer: CodeSSM (Bi-directional), CodeSSM-Uni (Uni-directional) and CodeF (Discrete Fourier Transform in place of SSM). Replacing SSM with DFT reduces the modeling capacity significantly.}
\label{bigs_bigf}
\end{center}
\vskip -0.2in
\end{figure}

To answer this question, we performed an ablation study by replacing the SSM transformation in CodeSSM with DFT and creating a unidirectional variant of CodeSSM. The results are presented in \cref{bigs_bigf} (and in \cref{MLM Eval}). We observe that both variants perform significantly worse compared to CodeSSM. However, it is important to note that CodeF performs better than CodeSSM-uni, i.e., the advantages of SSM over DFT on code modeling are only realized with bidirectional SSMs. 

\subsection{Is CodeSSM robust to overfitting on small datasets?}
We study the robustness of the CodeSSM model in two ways. First, we study the impact of adding dropout to the CodeSSM layer. Across all tasks, including the complexity prediction task, which has only 3.6k samples, the performance of the model decreases when dropout is added. This performance degradation suggests that CodeSSM does not overfit during training regardless of the size of the finetuning dataset. Second, we evaluated the model on the adversarial NLCodeSearch task, by which informative tokens such as function and variable names have been replaced with noninformative generic tokens. The CodeSSM model performs very well in this task even with minimal training. 

The reduction in performance with dropout and the performance of CodeSSM in the adversarial text-to-code search task demonstrate its \textbf{robust semantic understanding capabilities} on code understanding tasks.  

\subsection{When does CodeSSM fail and why?}\label{SSM_fail}

We discuss two possible scenarios where the performance of SSMs on code understanding tasks can be inferior to that of the transformers.

As discussed in \cref{modeling}, transformers can outperform SSMs as the number of samples increases. This phenomenon can hinder the performance of SSMs on some tasks such as Type Inference. This benchmark consists of over 9M type annotations, which is significantly larger than the samples of other tasks. The total number of predictions (hence, feedback to the model through loss) is even larger since there is one prediction for every token. However, when the number of samples is large and the context length of finetuning datasets is long, SSMs can surpass transformers in such scenarios (as seen in the SQA benchmark). 

SSMs can effectively capture long-range dependencies, but they do not model local dependencies \citep{spade} very well. The type inference benchmark requires the model to understand both short- and long-range dependencies. For example, in the code sample \texttt{x = y+1; z = func(x)}, the type of \texttt{x} can be inferred from the local context (variable \texttt{y}), while the type of \texttt{z} can only be inferred from the return type of the function \texttt{func}, which requires capturing long-range dependencies. We also observe that this problem becomes more challenging when the local context is complicated. We provide a qualitative example of the failure of CodeSSM in the type inference task in \cref{failurecase}. 
We can potentially improve CodeSSM's performance on this task by making the step size of SSMs (see \cref{background}) data dependent \cite{smr} or with multiple SSM kernels (\cref{failurecase}), which can improve its understanding of short-range dependencies.

\subsection{Expressivity gaps in SSMs and their fixes}
Some recent theoretical works document the expressivity limitations of SSMs. In particular, \citet{stateillusion} shows that SSMs cannot express some important problems that require state tracking. This class of problems includes evaluating Python code. However, it should be noted that transformers have also been shown to not express this class of problems while a single layer RNN can \cite{transformer-parallelism}. However, the expressive power of SSMs can be improved through different architectural improvements. For example, \citet{stateillusion} showed that making the input of the transition matrix independent or adding nonlinearities can extend the expressive power of SSM. Similarly, \citet{ssmdynamictoken} showed that when SSMs are preceded and followed by feed-forward layers, they have the same dynamic token abilities as transformers. 

Another limitation is the requirement of specific reparameterization. \citet{spectralssm} proposed a new SSM formulation by augmenting spectral filters with negative eigenvalues, which removed the requirement of reparameterization. The SSM variant used in CodeSSM is  S4D. S4D also has negative eigenvalues but still requires specific reparameterization. The best way to parametrize SSM and their practical benefits remain an open research question \cite{stablessm}.




\section{Related Work}
 Transformer models have been used in code intelligence tasks \cite{codebert, unixcoder, codet5, codet5p}. However, they struggle to scale with long sequence lengths due to the quadratic complexity of self-attention \cite{attentionisallyouneed}. Various approaches have been proposed to address quadratic complexity, including linear attention \cite{linear-attention} and methods using sparse attention patterns \cite{longt5, bigbird, lsg}. LongCoder \cite{longcoder} uses the sparse attention mechanism. 
 
 However, efficient transformers have seen limited practical adoption due to a minimal gain in computational cost in practice \cite{limitation-efficient-transformer} and a potential degradation in performance \cite{devil-in-linear-transformer}. \citet{flashattention} and \citet{flashattention-2} introduced hardware-level optimizations, but the benefits of these optimizations are hardware-specific and do not necessarily apply to newer or different hardware architectures \cite{flashattention-3}.   

Moreover, self-attention models are trained with a fixed context window and cannot be used beyond the pretraining context window. Techniques such as ALiBi \cite{alibi} and RoPE \cite{Rope} have been proposed to alleviate this problem. HiRoPE \cite{hirope} adapted RoPE to code models. However, these methods come with their own set of challenges \cite{lim-rope, lim-rope-fourier, lim-rope-mix}. 

In contrast, SSMs, and hence CodeSSM, are efficient by design and have been shown to better capture long-range dependencies compared to self-attention \cite{s4}. Additionally, CodeSSM does not have positional embedding, alleviating the issues of position-specific biases and length generalization. 

\citet{repo-1} and \citet{repo-3} improve base transformer models for longer contexts using repository-level information retrieval, while \citet{static-syntax, static-dataflow, static-callgraph, static-comp, static-1, repo-2} use static analysis tools for repository-level tasks. The usage of static analysis tools is complementary to CodeSSM, as the performance of CodeSSM can also be further improved by using these tools. Additionally, CodeSSM can be a drop-in replacement as an efficient retriever model. 

\section{Conclusions}
State-space models (SSMs) have emerged as an efficient alternative to transformers.
In this work, we investigated the effectiveness of SSM in code understanding tasks by introducing CodeSSM.
Our results show a significant potential for CodeSSM in code modeling and understanding. Our findings show that SSMs can be a robust and sample-efficient alternative to transformers, offering the added benefit of length extrapolation.

Although the training pipeline and architectural improvements for transformers have evolved extensively in the past decade,  the optimal training conditions for SSMs have yet to be thoroughly investigated. This work represents a step toward understanding these conditions and uncovering the advantages of SSMs. In particular, the performance gains achieved by CodeSSM were obtained under suboptimal conditions, indicating the potential for further enhancements. 
With advancements in training methodologies and architectural designs for SSM-based models, their performance could be significantly improved.

\section*{Limitations}
Although we have done an extensive study of SSMs for code understanding, our work has some limitations.
\begin{itemize}
   \item \textbf{Single kernel transformation:} We have used a simple implementation of S4d \cite{s4d} in the CodeSSM layer. This implementation has only one SSM kernel shared across all input channels. Incorporating multiple kernels into an SSM block can further improve performance. 
   \item \textbf{Short-range dependencies:} We have discussed SSMs' limitations in understanding short-range dependencies in the context of the type inference task. However, we do not provide a solution to this problem. While some methods have been proposed to alleviate this issue in the literature, we leave the their exploration in the context of code understanding for future work.  
   \item \textbf{Generative tasks:} CodeSSM currently uses SSMs only in convolution mode. Due to bi-directional SSMs, it cannot be used for auto-regressive tasks. As a result, it is not applicable to generative tasks. Hence, we cannot compare CodeSSM with Mamba \cite{mamba} or Codestral Mamba \footnote{\url{https://mistral.ai/news/codestral-mamba}}. To alleviate this limitation, future work can utilize the recurrent view of SSM to enable generative capabilities. 
\end{itemize}

\section*{Ethical Considerations}
We have used publicly existing datasets. CodeSearchNet dataset can contain personally identifiable information (PII). While the StarCoder dataset attempts to remove PII, there might still be some PII in the dataset.  It is not clear whether our model can leak such data if present in the pre-training dataset, but we acknowledge that such potential risk exists. In addition, our work involved training multiple models. Although the size of the model and the data set was small, training can still have environmental impacts.



\bibliography{custom}

\begin{thebibliography}{61}
\providecommand{\natexlab}[1]{#1}

\bibitem[{Agarwal et~al.(2024)Agarwal, Suo, Chen, and Hazan}]{spectralssm}
Naman Agarwal, Daniel Suo, Xinyi Chen, and Elad Hazan. 2024.
\newblock \href {https://arxiv.org/abs/2312.06837} {Spectral state space models}.
\newblock \emph{Preprint}, arXiv:2312.06837.

\bibitem[{Bi et~al.(2024)Bi, Wan, Wang, Zhang, Guan, Lu, Zhang, Sui, Jin, and Shi}]{static-comp}
Zhangqian Bi, Yao Wan, Zheng Wang, Hongyu Zhang, Batu Guan, Fangxin Lu, Zili Zhang, Yulei Sui, Hai Jin, and Xuanhua Shi. 2024.
\newblock \href {https://doi.org/10.18653/v1/2024.findings-acl.138} {Iterative refinement of project-level code context for precise code generation with compiler feedback}.
\newblock In \emph{Findings of the Association for Computational Linguistics: ACL 2024}, pages 2336--2353, Bangkok, Thailand.

\bibitem[{Chen et~al.(2023{\natexlab{a}})Chen, Ding, Alowain, Chen, and Wagner}]{diversevul}
Yizheng Chen, Zhoujie Ding, Lamya Alowain, Xinyun Chen, and David Wagner. 2023{\natexlab{a}}.
\newblock \href {https://arxiv.org/abs/2304.00409} {Diversevul: A new vulnerable source code dataset for deep learning based vulnerability detection}.
\newblock \emph{Preprint}, arXiv:2304.00409.

\bibitem[{Chen et~al.(2023{\natexlab{b}})Chen, Lv, Lin, Chen, Wu, Huang, Li, and Yan}]{lim-rope}
Yuhan Chen, Ang Lv, Ting-En Lin, Chang~Heng Chen, Yuchuan Wu, Fei Huang, Yongbin Li, and Rui Yan. 2023{\natexlab{b}}.
\newblock \href {https://api.semanticscholar.org/CorpusID:266053571} {Fortify the shortest stave in attention: Enhancing context awareness of large language models for effective tool use}.
\newblock In \emph{Annual Meeting of the Association for Computational Linguistics}.

\bibitem[{Cheng et~al.(2024)Cheng, Wu, and Hu}]{static-dataflow}
Wei Cheng, Yuhan Wu, and Wei Hu. 2024.
\newblock \href {https://doi.org/10.18653/v1/2024.acl-long.431} {Dataflow-guided retrieval augmentation for repository-level code completion}.
\newblock In \emph{Proceedings of the 62nd Annual Meeting of the Association for Computational Linguistics (Volume 1: Long Papers)}, pages 7957--7977, Bangkok, Thailand.

\bibitem[{Clement et~al.(2021)Clement, Lu, Liu, Tufano, Drain, Duan, Sundaresan, and Svyatkovskiy}]{static-syntax}
Colin Clement, Shuai Lu, Xiaoyu Liu, Michele Tufano, Dawn Drain, Nan Duan, Neel Sundaresan, and Alexey Svyatkovskiy. 2021.
\newblock \href {https://doi.org/10.18653/v1/2021.emnlp-main.387} {Long-range modeling of source code files with e{WASH}: Extended window access by syntax hierarchy}.
\newblock In \emph{Proceedings of the 2021 Conference on Empirical Methods in Natural Language Processing}, pages 4713--4722, Online and Punta Cana, Dominican Republic.

\bibitem[{Condevaux and Harispe(2022)}]{lsg}
Charles Condevaux and S{\'e}bastien Harispe. 2022.
\newblock \href {https://api.semanticscholar.org/CorpusID:253157377} {Lsg attention: Extrapolation of pretrained transformers to long sequences}.
\newblock In \emph{Pacific-Asia Conference on Knowledge Discovery and Data Mining}.

\bibitem[{Dao(2024)}]{flashattention-2}
Tri Dao. 2024.
\newblock \href {https://openreview.net/forum?id=mZn2Xyh9Ec} {Flashattention-2: Faster attention with better parallelism and work partitioning}.
\newblock In \emph{The Twelfth International Conference on Learning Representations}.

\bibitem[{Dao et~al.(2022)Dao, Fu, Ermon, Rudra, and R\'{e}}]{flashattention}
Tri Dao, Dan Fu, Stefano Ermon, Atri Rudra, and Christopher R\'{e}. 2022.
\newblock \href {https://openreview.net/forum?id=H4DqfPSibmx} {Flashattention: Fast and memory-efficient exact attention with io-awareness}.
\newblock In \emph{Advances in Neural Information Processing Systems}, volume~35, pages 16344--16359. Curran Associates, Inc.

\bibitem[{Devlin et~al.(2019)Devlin, Chang, Lee, and Toutanova}]{bert}
Jacob Devlin, Ming-Wei Chang, Kenton Lee, and Kristina Toutanova. 2019.
\newblock \href {https://doi.org/10.18653/v1/N19-1423} {{BERT}: Pre-training of deep bidirectional transformers for language understanding}.
\newblock In \emph{Proceedings of the 2019 Conference of the North {A}merican Chapter of the Association for Computational Linguistics: Human Language Technologies, Volume 1 (Long and Short Papers)}, pages 4171--4186, Minneapolis, Minnesota. Association for Computational Linguistics.

\bibitem[{Feng et~al.(2020)Feng, Guo, Tang, Duan, Feng, Gong, Shou, Qin, Liu, Jiang, and Zhou}]{codebert}
Zhangyin Feng, Daya Guo, Duyu Tang, Nan Duan, Xiaocheng Feng, Ming Gong, Linjun Shou, Bing Qin, Ting Liu, Daxin Jiang, and Ming Zhou. 2020.
\newblock \href {https://doi.org/10.18653/v1/2020.findings-emnlp.139} {{C}ode{BERT}: A pre-trained model for programming and natural languages}.
\newblock In \emph{Findings of the Association for Computational Linguistics: EMNLP 2020}, pages 1536--1547, Online. Association for Computational Linguistics.

\bibitem[{Glorioso et~al.(2024)Glorioso, Anthony, Tokpanov, Golubeva, Shyam, Whittington, Pilault, and Millidge}]{zamba2}
Paolo Glorioso, Quentin Anthony, Yury Tokpanov, Anna Golubeva, Vasudev Shyam, James Whittington, Jonathan Pilault, and Beren Millidge. 2024.
\newblock \href {https://arxiv.org/abs/2411.15242} {The zamba2 suite: Technical report}.
\newblock \emph{Preprint}, arXiv:2411.15242.

\bibitem[{Goel et~al.(2025)Goel, Lee, and Ramchandran}]{positionalbias}
Samarth Goel, Reagan~J. Lee, and Kannan Ramchandran. 2025.
\newblock \href {https://arxiv.org/abs/2412.15241} {Quantifying positional biases in text embedding models}.
\newblock \emph{Preprint}, arXiv:2412.15241.

\bibitem[{Gu and Dao(2024)}]{mamba}
Albert Gu and Tri Dao. 2024.
\newblock \href {https://openreview.net/forum?id=tEYskw1VY2} {Mamba: Linear-time sequence modeling with selective state spaces}.
\newblock In \emph{First Conference on Language Modeling}.

\bibitem[{Gu et~al.(2022{\natexlab{a}})Gu, Goel, Gupta, and R\'{e}}]{s4d}
Albert Gu, Karan Goel, Ankit Gupta, and Christopher R\'{e}. 2022{\natexlab{a}}.
\newblock \href {https://openreview.net/forum?id=yJE7iQSAep} {On the parameterization and initialization of diagonal state space models}.
\newblock In \emph{Advances in Neural Information Processing Systems}, volume~35, pages 35971--35983. Curran Associates, Inc.

\bibitem[{Gu et~al.(2022{\natexlab{b}})Gu, Goel, and Re}]{s4}
Albert Gu, Karan Goel, and Christopher Re. 2022{\natexlab{b}}.
\newblock \href {https://openreview.net/forum?id=uYLFoz1vlAC} {Efficiently modeling long sequences with structured state spaces}.
\newblock In \emph{International Conference on Learning Representations}.

\bibitem[{Gu et~al.(2021)Gu, Johnson, Goel, Saab, Dao, Rudra, and R\'{e}}]{linear-ssm}
Albert Gu, Isys Johnson, Karan Goel, Khaled Saab, Tri Dao, Atri Rudra, and Christopher R\'{e}. 2021.
\newblock \href {https://proceedings.neurips.cc/paper_files/paper/2021/file/05546b0e38ab9175cd905eebcc6ebb76-Paper.pdf} {Combining recurrent, convolutional, and continuous-time models with linear state space layers}.
\newblock In \emph{Advances in Neural Information Processing Systems}, volume~34, pages 572--585.

\bibitem[{Guan et~al.(2024)Guan, Liu, Liu, Peng, Liu, Sun, Jiang, Li, Liu, and Zhu}]{repo-1}
Zhanming Guan, Junlin Liu, Jierui Liu, Chao Peng, Dexin Liu, Ningyuan Sun, Bo~Jiang, Wenchao Li, Jie Liu, and Hang Zhu. 2024.
\newblock \href {https://arxiv.org/abs/2412.08063} {Contextmodule: Improving code completion via repository-level contextual information}.
\newblock \emph{Preprint}, arXiv:2412.08063.

\bibitem[{Guo et~al.(2022{\natexlab{a}})Guo, Lu, Duan, Wang, Zhou, and Yin}]{unixcoder}
Daya Guo, Shuai Lu, Nan Duan, Yanlin Wang, Ming Zhou, and Jian Yin. 2022{\natexlab{a}}.
\newblock \href {https://doi.org/10.18653/v1/2022.acl-long.499} {{U}ni{X}coder: Unified cross-modal pre-training for code representation}.
\newblock In \emph{Proceedings of the 60th Annual Meeting of the Association for Computational Linguistics (Volume 1: Long Papers)}, pages 7212--7225, Dublin, Ireland. Association for Computational Linguistics.

\bibitem[{Guo et~al.(2023)Guo, Xu, Duan, Yin, and McAuley}]{longcoder}
Daya Guo, Canwen Xu, Nan Duan, Jian Yin, and Julian McAuley. 2023.
\newblock \href {https://arxiv.org/pdf/2306.14893} {Longcoder: A long-range pre-trained language model for code completion}.
\newblock In \emph{International Conference on Machine Learning}.

\bibitem[{Guo et~al.(2022{\natexlab{b}})Guo, Ainslie, Uthus, Ontanon, Ni, Sung, and Yang}]{longt5}
Mandy Guo, Joshua Ainslie, David Uthus, Santiago Ontanon, Jianmo Ni, Yun-Hsuan Sung, and Yinfei Yang. 2022{\natexlab{b}}.
\newblock \href {https://doi.org/10.18653/v1/2022.findings-naacl.55} {{L}ong{T}5: {E}fficient text-to-text transformer for long sequences}.
\newblock In \emph{Findings of the Association for Computational Linguistics: NAACL 2022}, pages 724--736, Seattle, United States. Association for Computational Linguistics.

\bibitem[{Hendrycks and Gimpel(2016)}]{GELU}
Dan Hendrycks and Kevin Gimpel. 2016.
\newblock \href {https://api.semanticscholar.org/CorpusID:125617073} {Gaussian error linear units (gelus)}.
\newblock \emph{arXiv: Learning}.

\bibitem[{Hua et~al.(2025{\natexlab{a}})Hua, Jiang, Lv, Zhang, Ding, Sun, Qi, Fan, Zhu, and Zhou}]{lim-rope-fourier}
Ermo Hua, Che Jiang, Xingtai Lv, Kaiyan Zhang, Ning Ding, Youbang Sun, Biqing Qi, Yuchen Fan, Xuekai Zhu, and Bowen Zhou. 2025{\natexlab{a}}.
\newblock \href {https://arxiv.org/abs/2412.17739} {Fourier position embedding: Enhancing attention's periodic extension for length generalization}.
\newblock \emph{Preprint}, arXiv:2412.17739.

\bibitem[{Hua et~al.(2025{\natexlab{b}})Hua, Jiang, Lv, Zhang, Ding, Sun, Qi, Fan, Zhu, and Zhou}]{FoPE}
Ermo Hua, Che Jiang, Xingtai Lv, Kaiyan Zhang, Ning Ding, Youbang Sun, Biqing Qi, Yuchen Fan, Xuekai Zhu, and Bowen Zhou. 2025{\natexlab{b}}.
\newblock \href {https://arxiv.org/abs/2412.17739} {Fourier position embedding: Enhancing attention's periodic extension for length generalization}.
\newblock \emph{Preprint}, arXiv:2412.17739.

\bibitem[{Hua et~al.(2022)Hua, Dai, Liu, and Le}]{huaetal}
Weizhe Hua, Zihang Dai, Hanxiao Liu, and Quoc Le. 2022.
\newblock \href {https://proceedings.mlr.press/v162/hua22a.html} {Transformer quality in linear time}.
\newblock In \emph{Proceedings of the 39th International Conference on Machine Learning}, volume 162 of \emph{Proceedings of Machine Learning Research}, pages 9099--9117. PMLR.

\bibitem[{Husain et~al.(2019)Husain, Wu, Gazit, Allamanis, and Brockschmidt}]{CodeSearchNet}
Hamel Husain, Hongqiu Wu, Tiferet Gazit, Miltiadis Allamanis, and Marc Brockschmidt. 2019.
\newblock \href {https://api.semanticscholar.org/CorpusID:202712680} {Codesearchnet challenge: Evaluating the state of semantic code search}.
\newblock \emph{ArXiv}, abs/1909.09436.

\bibitem[{Jeon et~al.(2023)Jeon, yeop Baik, Hahn, Han, and Ko}]{complexity}
Mingi Jeon, Seung yeop Baik, Joonghyuk Hahn, Yo-Sub Han, and Sang-Ki Ko. 2023.
\newblock \href {https://openreview.net/forum?id=9irBKvxsw9} {Deep learning-based source code complexity prediction}.

\bibitem[{Jesse and Devanbu(2022)}]{Type-inference}
Kevin Jesse and Premkumar~T. Devanbu. 2022.
\newblock \href {https://doi.org/10.1145/3524842.3528507} {Manytypes4typescript: a comprehensive typescript dataset for sequence-based type inference}.
\newblock In \emph{Proceedings of the 19th International Conference on Mining Software Repositories}, MSR '22, page 294–298, New York, NY, USA. Association for Computing Machinery.

\bibitem[{Katharopoulos et~al.(2020)Katharopoulos, Vyas, Pappas, and Fleuret}]{linear-attention}
Angelos Katharopoulos, Apoorv Vyas, Nikolaos Pappas, and Fran\c{c}ois Fleuret. 2020.
\newblock \href {http://proceedings.mlr.press/v119/katharopoulos20a/katharopoulos20a.pdf} {Transformers are rnns: fast autoregressive transformers with linear attention}.
\newblock In \emph{Proceedings of the 37th International Conference on Machine Learning}, ICML'20. JMLR.org.

\bibitem[{Le~Bronnec et~al.(2024)Le~Bronnec, Duong, Ravaut, Allauzen, Chen, Guigue, Lumbreras, Soulier, and Gallinari}]{locost}
Florian Le~Bronnec, Song Duong, Mathieu Ravaut, Alexandre Allauzen, Nancy Chen, Vincent Guigue, Alberto Lumbreras, Laure Soulier, and Patrick Gallinari. 2024.
\newblock \href {https://aclanthology.org/2024.eacl-long.69/} {{LOCOST}: State-space models for long document abstractive summarization}.
\newblock In \emph{Proceedings of the 18th Conference of the European Chapter of the Association for Computational Linguistics (Volume 1: Long Papers)}, pages 1144--1159, St. Julian{'}s, Malta. Association for Computational Linguistics.

\bibitem[{Lee-Thorp et~al.(2022)Lee-Thorp, Ainslie, Eckstein, and Ontanon}]{fnet}
James Lee-Thorp, Joshua Ainslie, Ilya Eckstein, and Santiago Ontanon. 2022.
\newblock \href {https://doi.org/10.18653/v1/2022.naacl-main.319} {{FN}et: Mixing tokens with {F}ourier transforms}.
\newblock In \emph{Proceedings of the 2022 Conference of the North American Chapter of the Association for Computational Linguistics: Human Language Technologies}, pages 4296--4313, Seattle, United States. Association for Computational Linguistics.

\bibitem[{Li et~al.(2023)Li, allal, Zi, Muennighoff, Kocetkov, Mou, Marone, Akiki, LI, Chim, Liu, Zheltonozhskii, Zhuo, Wang, Dehaene, Lamy-Poirier, Monteiro, Gontier, Yee, Umapathi, Zhu, Lipkin, Oblokulov, Wang, Murthy, Stillerman, Patel, Abulkhanov, Zocca, Dey, Zhang, Bhattacharyya, Yu, Luccioni, Villegas, Zhdanov, Lee, Timor, Ding, Schlesinger, Schoelkopf, Ebert, Dao, Mishra, Gu, Anderson, Dolan-Gavitt, Contractor, Reddy, Fried, Bahdanau, Jernite, Ferrandis, Hughes, Wolf, Guha, Werra, and de~Vries}]{starcoder}
Raymond Li, Loubna~Ben allal, Yangtian Zi, Niklas Muennighoff, Denis Kocetkov, Chenghao Mou, Marc Marone, Christopher Akiki, Jia LI, Jenny Chim, Qian Liu, Evgenii Zheltonozhskii, Terry~Yue Zhuo, Thomas Wang, Olivier Dehaene, Joel Lamy-Poirier, Joao Monteiro, Nicolas Gontier, Ming-Ho Yee, and 39 others. 2023.
\newblock \href {https://openreview.net/forum?id=KoFOg41haE} {Starcoder: may the source be with you!}
\newblock \emph{Transactions on Machine Learning Research}.

\bibitem[{Lin et~al.(2024)Lin, Lv, Chen, Zhu, Song, Zhu, and Yan}]{lim-rope-mix}
Hongzhan Lin, Ang Lv, Yuhan Chen, Chen Zhu, Yang Song, Hengshu Zhu, and Rui Yan. 2024.
\newblock \href {https://openreview.net/forum?id=RcPHbofiCN} {Mixture of in-context experts enhance {LLM}s' long context awareness}.
\newblock In \emph{The Thirty-eighth Annual Conference on Neural Information Processing Systems}.

\bibitem[{Liu et~al.(2024{\natexlab{a}})Liu, Chen, Liu, Peng, and Lou}]{static-1}
Junwei Liu, Yixuan Chen, Mingwei Liu, Xin Peng, and Yiling Lou. 2024{\natexlab{a}}.
\newblock \href {https://arxiv.org/abs/2406.10018} {Stall+: Boosting llm-based repository-level code completion with static analysis}.
\newblock \emph{Preprint}, arXiv:2406.10018.

\bibitem[{Liu et~al.(2024{\natexlab{b}})Liu, Lin, Hewitt, Paranjape, Bevilacqua, Petroni, and Liang}]{lost-in-the-middle}
Nelson~F. Liu, Kevin Lin, John Hewitt, Ashwin Paranjape, Michele Bevilacqua, Fabio Petroni, and Percy Liang. 2024{\natexlab{b}}.
\newblock \href {https://doi.org/10.1162/tacl_a_00638} {Lost in the middle: How language models use long contexts}.
\newblock \emph{Transactions of the Association for Computational Linguistics}, 12:157--173.

\bibitem[{Lomshakov et~al.(2024)Lomshakov, Podivilov, Savin, Baryshnikov, Lisevych, and Nikolenko}]{static-callgraph}
Vadim Lomshakov, Andrey Podivilov, Sergey Savin, Oleg Baryshnikov, Alena Lisevych, and Sergey Nikolenko. 2024.
\newblock \href {https://doi.org/10.18653/v1/2024.emnlp-industry.65} {{P}ro{C}on{S}u{L}: Project context for code summarization with {LLM}s}.
\newblock In \emph{Proceedings of the 2024 Conference on Empirical Methods in Natural Language Processing: Industry Track}, pages 866--880, Miami, Florida, US.

\bibitem[{Lu et~al.(2021)Lu, Guo, Ren, Huang, Svyatkovskiy, Blanco, Clement, Drain, Jiang, Tang, Li, Zhou, Shou, Zhou, Tufano, GONG, Zhou, Duan, Sundaresan, Deng, Fu, and LIU}]{codexglue}
Shuai Lu, Daya Guo, Shuo Ren, Junjie Huang, Alexey Svyatkovskiy, Ambrosio Blanco, Colin Clement, Dawn Drain, Daxin Jiang, Duyu Tang, Ge~Li, Lidong Zhou, Linjun Shou, Long Zhou, Michele Tufano, MING GONG, Ming Zhou, Nan Duan, Neel Sundaresan, and 3 others. 2021.
\newblock \href {https://openreview.net/forum?id=6lE4dQXaUcb} {Code{XGLUE}: A machine learning benchmark dataset for code understanding and generation}.
\newblock In \emph{Thirty-fifth Conference on Neural Information Processing Systems Datasets and Benchmarks Track (Round 1)}.

\bibitem[{Ma et~al.(2024)Ma, Yang, Cao, Li, Huang, and Li}]{repo-2}
Yingwei Ma, Qingping Yang, Rongyu Cao, Binhua Li, Fei Huang, and Yongbin Li. 2024.
\newblock \href {https://arxiv.org/abs/2406.01422} {How to understand whole software repository?}
\newblock \emph{Preprint}, arXiv:2406.01422.

\bibitem[{Merrill et~al.(2024)Merrill, Petty, and Sabharwal}]{stateillusion}
William Merrill, Jackson Petty, and Ashish Sabharwal. 2024.
\newblock The illusion of state in state-space models.
\newblock In \emph{Proceedings of the 41st International Conference on Machine Learning}, ICML'24. JMLR.org.

\bibitem[{Merrill and Sabharwal(2023)}]{transformer-parallelism}
William Merrill and Ashish Sabharwal. 2023.
\newblock \href {https://doi.org/10.1162/tacl_a_00562} {The parallelism tradeoff: Limitations of log-precision transformers}.
\newblock \emph{Transactions of the Association for Computational Linguistics}, 11:531--545.

\bibitem[{Nishikawa and Suzuki(2025)}]{ssmdynamictoken}
Naoki Nishikawa and Taiji Suzuki. 2025.
\newblock \href {https://openreview.net/forum?id=QFgbJOYJSE} {State space models are provably comparable to transformers in dynamic token selection}.
\newblock In \emph{The Thirteenth International Conference on Learning Representations}.

\bibitem[{Peng et~al.(2024)Peng, Quesnelle, Fan, and Shippole}]{yarn}
Bowen Peng, Jeffrey Quesnelle, Honglu Fan, and Enrico Shippole. 2024.
\newblock \href {https://openreview.net/forum?id=wHBfxhZu1u} {Ya{RN}: Efficient context window extension of large language models}.
\newblock In \emph{The Twelfth International Conference on Learning Representations}.

\bibitem[{Press et~al.(2022)Press, Smith, and Lewis}]{alibi}
Ofir Press, Noah Smith, and Mike Lewis. 2022.
\newblock \href {https://openreview.net/forum?id=R8sQPpGCv0} {Train short, test long: Attention with linear biases enables input length extrapolation}.
\newblock In \emph{International Conference on Learning Representations}.

\bibitem[{Qi et~al.(2024)Qi, Gao, Zhang, Li, Liu, Wu, and Zhou}]{smr}
Biqing Qi, Junqi Gao, Kaiyan Zhang, Dong Li, Jianxing Liu, Ligang Wu, and Bowen Zhou. 2024.
\newblock \href {https://doi.org/10.18653/v1/2024.findings-acl.483} {{SMR}: State memory replay for long sequence modeling}.
\newblock In \emph{Findings of the Association for Computational Linguistics: ACL 2024}, pages 8102--8116, Bangkok, Thailand. Association for Computational Linguistics.

\bibitem[{Qin et~al.(2022)Qin, Han, Sun, Li, Kong, Barnes, and Zhong}]{devil-in-linear-transformer}
Zhen Qin, Xiaodong Han, Weixuan Sun, Dongxu Li, Lingpeng Kong, Nick Barnes, and Yiran Zhong. 2022.
\newblock \href {https://doi.org/10.18653/v1/2022.emnlp-main.473} {The devil in linear transformer}.
\newblock In \emph{Proceedings of the 2022 Conference on Empirical Methods in Natural Language Processing}, pages 7025--7041, Abu Dhabi, United Arab Emirates. Association for Computational Linguistics.

\bibitem[{Shah et~al.(2024)Shah, Bikshandi, Zhang, Thakkar, Ramani, and Dao}]{flashattention-3}
Jay Shah, Ganesh Bikshandi, Ying Zhang, Vijay Thakkar, Pradeep Ramani, and Tri Dao. 2024.
\newblock \href {https://openreview.net/forum?id=tVConYid20} {Flashattention-3: Fast and accurate attention with asynchrony and low-precision}.
\newblock In \emph{The Thirty-eighth Annual Conference on Neural Information Processing Systems}.

\bibitem[{Srivastava et~al.(2014)Srivastava, Hinton, Krizhevsky, Sutskever, and Salakhutdinov}]{dropout}
Nitish Srivastava, Geoffrey Hinton, Alex Krizhevsky, Ilya Sutskever, and Ruslan Salakhutdinov. 2014.
\newblock \href {http://jmlr.org/papers/v15/srivastava14a.html} {Dropout: a simple way to prevent neural networks from overfitting}.
\newblock \emph{J. Mach. Learn. Res.}, 15(1):1929–1958.

\bibitem[{Su et~al.(2024)Su, Ahmed, Lu, Pan, Bo, and Liu}]{Rope}
Jianlin Su, Murtadha Ahmed, Yu~Lu, Shengfeng Pan, Wen Bo, and Yunfeng Liu. 2024.
\newblock \href {https://doi.org/10.1016/j.neucom.2023.127063} {Roformer: Enhanced transformer with rotary position embedding}.
\newblock \emph{Neurocomput.}, 568(C).

\bibitem[{Tay et~al.(2021)Tay, Dehghani, Gupta, Aribandi, Bahri, Qin, and Metzler}]{pre-trained-conv}
Yi~Tay, Mostafa Dehghani, Jai~Prakash Gupta, Vamsi Aribandi, Dara Bahri, Zhen Qin, and Donald Metzler. 2021.
\newblock \href {https://doi.org/10.18653/v1/2021.acl-long.335} {Are pretrained convolutions better than pretrained transformers?}
\newblock In \emph{Proceedings of the 59th Annual Meeting of the Association for Computational Linguistics and the 11th International Joint Conference on Natural Language Processing (Volume 1: Long Papers)}, pages 4349--4359, Online. Association for Computational Linguistics.

\bibitem[{Vaswani et~al.(2017)Vaswani, Shazeer, Parmar, Uszkoreit, Jones, Gomez, Kaiser, and Polosukhin}]{attentionisallyouneed}
Ashish Vaswani, Noam Shazeer, Niki Parmar, Jakob Uszkoreit, Llion Jones, Aidan~N Gomez, \L~ukasz Kaiser, and Illia Polosukhin. 2017.
\newblock \href {https://proceedings.neurips.cc/paper_files/paper/2017/file/3f5ee243547dee91fbd053c1c4a845aa-Paper.pdf} {Attention is all you need}.
\newblock In \emph{Advances in Neural Information Processing Systems}, volume~30. Curran Associates, Inc.

\bibitem[{Wang et~al.(2024)Wang, He, and Chen}]{repo-3}
Jicheng Wang, Yifeng He, and Hao Chen. 2024.
\newblock \href {https://arxiv.org/abs/2409.13122} {Repogenreflex: Enhancing repository-level code completion with verbal reinforcement and retrieval-augmented generation}.
\newblock \emph{Preprint}, arXiv:2409.13122.

\bibitem[{Wang et~al.(2023{\natexlab{a}})Wang, Yan, Gu, and Rush}]{bigs}
Junxiong Wang, Jing~Nathan Yan, Albert Gu, and Alexander Rush. 2023{\natexlab{a}}.
\newblock \href {https://doi.org/10.18653/v1/2023.findings-emnlp.5} {Pretraining without attention}.
\newblock In \emph{Findings of the Association for Computational Linguistics: EMNLP 2023}, pages 58--69, Singapore. Association for Computational Linguistics.

\bibitem[{Wang and Li(2024)}]{stablessm}
Shida Wang and Qianxiao Li. 2024.
\newblock Stablessm: alleviating the curse of memory in state-space models through stable reparameterization.
\newblock In \emph{Proceedings of the 41st International Conference on Machine Learning}, ICML'24.

\bibitem[{Wang et~al.(2020)Wang, Li, Ma, Xia, and Jin}]{clone}
Wenhan Wang, Ge~Li, Bo~Ma, Xin Xia, and Zhi Jin. 2020.
\newblock Detecting code clones with graph neural network and flow-augmented abstract syntax tree.
\newblock In \emph{2020 IEEE 27th International Conference on Software Analysis, Evolution and Reengineering (SANER)}, pages 261--271. IEEE.

\bibitem[{Wang et~al.(2023{\natexlab{b}})Wang, Le, Gotmare, Bui, Li, and Hoi}]{codet5p}
Yue Wang, Hung Le, Akhilesh~Deepak Gotmare, Nghi D.~Q. Bui, Junnan Li, and Steven C.~H. Hoi. 2023{\natexlab{b}}.
\newblock \href {https://arxiv.org/abs/2305.07922} {Codet5+: Open code large language models for code understanding and generation}.
\newblock \emph{Preprint}, arXiv:2305.07922.

\bibitem[{Wang et~al.(2021)Wang, Wang, Joty, and Hoi}]{codet5}
Yue Wang, Weishi Wang, Shafiq Joty, and Steven~C.H. Hoi. 2021.
\newblock \href {https://doi.org/10.18653/v1/2021.emnlp-main.685} {{C}ode{T}5: Identifier-aware unified pre-trained encoder-decoder models for code understanding and generation}.
\newblock In \emph{Proceedings of the 2021 Conference on Empirical Methods in Natural Language Processing}, pages 8696--8708, Online and Punta Cana, Dominican Republic. Association for Computational Linguistics.

\bibitem[{Yang et~al.(2025)Yang, Ackermann, He, Feng, Zhang, Feng, Ye, He, and Wang}]{limitation-efficient-transformer}
Kai Yang, Jan Ackermann, Zhenyu He, Guhao Feng, Bohang Zhang, Yunzhen Feng, Qiwei Ye, Di~He, and Liwei Wang. 2025.
\newblock \href {https://openreview.net/pdf?id=xLikRS9OhW} {Do efficient transformers really save computation?}
\newblock In \emph{Proceedings of the 41st International Conference on Machine Learning}, ICML'24. JMLR.org.

\bibitem[{Zaheer et~al.(2020)Zaheer, Guruganesh, Dubey, Ainslie, Alberti, Ontanon, Pham, Ravula, Wang, Yang, and Ahmed}]{bigbird}
Manzil Zaheer, Guru Guruganesh, Kumar~Avinava Dubey, Joshua Ainslie, Chris Alberti, Santiago Ontanon, Philip Pham, Anirudh Ravula, Qifan Wang, Li~Yang, and Amr Ahmed. 2020.
\newblock \href {https://proceedings.neurips.cc/paper_files/paper/2020/file/c8512d142a2d849725f31a9a7a361ab9-Paper.pdf} {Big bird: Transformers for longer sequences}.
\newblock In \emph{Advances in Neural Information Processing Systems}, volume~33, pages 17283--17297. Curran Associates, Inc.

\bibitem[{Zhang et~al.(2024)Zhang, Li, Zhang, and Jin}]{hirope}
Kechi Zhang, Ge~Li, Huangzhao Zhang, and Zhi Jin. 2024.
\newblock \href {https://doi.org/10.18653/v1/2024.acl-long.735} {{H}i{R}o{PE}: Length extrapolation for code models using hierarchical position}.
\newblock In \emph{Proceedings of the 62nd Annual Meeting of the Association for Computational Linguistics (Volume 1: Long Papers)}, pages 13615--13627, Bangkok, Thailand.

\bibitem[{Zhou et~al.(2019)Zhou, Liu, Siow, Du, and Liu}]{devign}
Yaqin Zhou, Shangqing Liu, Jingkai Siow, Xiaoning Du, and Yang Liu. 2019.
\newblock Devign: Effective vulnerability identification by learning comprehensive program semantics via graph neural networks.
\newblock In \emph{Advances in Neural Information Processing Systems}, pages 10197--10207.

\bibitem[{Zuo et~al.(2024)Zuo, Liu, Jiao, Charles, Manavoglu, Zhao, and Gao}]{spade}
Simiao Zuo, Xiaodong Liu, Jian Jiao, Denis~X Charles, Eren Manavoglu, Tuo Zhao, and Jianfeng Gao. 2024.
\newblock \href {https://openreview.net/forum?id=uUIFTjBREk} {Efficient hybrid long sequence modeling with state space augmented transformers}.
\newblock In \emph{First Conference on Language Modeling}.

\end{thebibliography}

\appendix


\section{Background} \label{background}
\subsection{State Space Model}
A state space model (SSM for short) 
maps a 1-dimensional input signal $u(t)$ to an N-dimensional latent state $x(t)$ and then projects it back to a 1-dimensional output. The transformations follow the equations
\begin{equation} \label{ssm-eq}
\begin{split}
    x'(t) &= \mathbf{A}x(t) + \mathbf{B}u(t) \\
    y(t) &= \mathbf{C}x(t) + \mathbf{D}u(t)
\end{split}
\end{equation}
where $\mathbf{A}$ controls how the system evolves, and $\mathbf{B}$, $\mathbf{C}$, $\mathbf{D}$ are projection parameters. 
Previous work \cite{linear-ssm, s4, s4d} considered SSM as a black-box representation within a deep sequence model. In this context, the parameters $\mathbf{A}$, $\mathbf{B}$, $\mathbf{C}$, $\mathbf{D}$ are learned with gradient descent.  For simplicity, we can assume $\mathbf{D} = 0$, as it corresponds to a skip connection and is straightforward to compute. To apply these models to discrete input sequences, Eq. \ref{ssm-eq} is discretized with a step size of $\Delta$.

\begin{equation} \label{dssm-eq}
    \begin{split}
        x_k &= \overline{\mathbf{A}}x_{k-1} + \overline{\mathbf{B}}u_k \\
        y_k &= \overline{\mathbf{C}}x_k 
    \end{split}
\end{equation}

Eq. \ref{dssm-eq} can be interpreted as a convolution between the input sequence and a convolution kernel, which is a function of $\mathbf{A}$, $\mathbf{B}$ and $\mathbf{C}$. This convolution view of SSM is expressed as follows :
\begin{equation}
    \begin{split}
        y_k &= \overline{\mathbf{C}}\overline{\mathbf{A}}^{k}\overline{\mathbf{B}}u_0 + \overline{\mathbf{C}}\overline{\mathbf{A}}^{k-1}\overline{\mathbf{B}}u_1 + ... + \\ &\overline{\mathbf{C}}\overline{\mathbf{A}}\overline{\mathbf{B}}u_{k-1} + \overline{\mathbf{C}}\overline{\mathbf{B}}u_k \\
        y &= \overline{\mathbf{K}} * u \\
         \overline{\mathbf{K}} &\in \mathbb{R} \coloneqq K_L(\overline{\mathbf{A}},\overline{\mathbf{B}},\overline{\mathbf{C}}) 
    \end{split}
\end{equation}

where $\overline{\mathbf{K}}$ is the SSM convolution kernel. 

Throughout the paper, we denote the input length as $L$ and the hidden dimension as $d$. We follow \cite{s4d} for all the parameterization.

\subsection{BiGS Model}
BiGS model \cite{bigs} is a bidirectional gated architecture which consists of two SSM layers as shown in Fig. \ref{bigs_arch}. This gated architecture is adapted from the gated attention unit described in \cite{huaetal}. 

A single layer of the BiGS model has a three stage computation. 
The first stage utilizes a feed-forward layer alongside gating, employing the GELU activation function \cite{GELU}. The second stage consists of two sequential SSM blocks with a multiplicative gating.
The third stage again uses feed-forward and gating. 

\begin{equation} \label{bigs1}
\begin{split}
   \mathbf{X} &= \text{LayerNorm}(\mathbf{X}_i)  \quad \in \mathbb{R}^{L\times d} \\
\mathbf{V} &= \sigma(\mathbf{W}_v\mathbf{X})  \quad \in \mathbb{R}^{L\times 3d} \\
\mathbf{F} &= \sigma(\mathbf{W}_f\mathbf{X})  \quad \in  \mathbb{R}^{L\times d}  \\
\mathbf{B} &= \sigma(\mathbf{W}_b\text{Flip}(\mathbf{X}))  \quad \in  \mathbb{R}^{L\times d} 
\end{split}
\end{equation}

\begin{equation} \label{bigs2}
\begin{split}
   \mathbf{U}_1 &= \mathbf{W}_{u_1}\text{SSM}(\mathbf{F}) \quad \in \mathbb{R}^{L\times d} \\
\mathbf{U}_2 &= \mathbf{W}_{u_2}\text{SSM}(\mathbf{B})  \quad \in \mathbb{R}^{L\times d} \\
\mathbf{U} &= \sigma(\mathbf{W}_u(\mathbf{U}_1 \otimes \text{Flip}(\mathbf{U}_2)))  \quad \in  \mathbb{R}^{L\times 3d}  
\end{split}
\end{equation}

\begin{equation} \label{bigs3}
\begin{split}
   \mathbf{O} &= \mathbf{W}_o (\mathbf{U} \otimes \mathbf{V}) \quad \in \mathbb{R}^{L\times d} \\
   \mathbf{X}_{i+1} &= \mathbf{O} + \mathbf{X}_i  \quad \in \mathbb{R}^{L\times d} 
\end{split}
\end{equation}

The BiGS model consists of 23 layers and includes a positional embedding.

\subsection{Model Size}
All the models we used, CodeSSM, BertCoder and RoCoder, has 190M parameters.

\section{Hardware Details}
We run all our experiments on Nvidia A100 GPUs with 80GB memory. All pretraining is done on 4 GPUs and all finetuning is done on a single GPU.

\section{Dataset}
\textbf{CodeSearchNet Dataset.} The CodeSearchNet dataset \citep{CodeSearchNet} used in this work is the unofficial dataset available on Huggingface hub \footnote{\url{https://huggingface.co/datasets/code-search-net/code_search_net}} using datasets library \footnote{\url{https://github.com/huggingface/datasets}}, since the official dataset is no longer accessible. The unofficial dataset is significantly smaller compared to the original CSN dataset used for training other attention-based models for code understanding, such as CodeBERT.

\textbf{StarCoder Dataset.} The StarCoder Dataset \citep{starcoder} is a cleaned and de-duplicated dataset with Personally Identifiable Information (PII) redacted. The dataset contains \texttt{git-issues} and \texttt{git-commits} along with 86 programming languages. Additionally, the authors add additional information to every sample such as GitHub star, repository name and path. We remove these information along with the accompanying tokens such as \texttt{<gh-stars>} before using the dataset.
To create the subset used in our work, we simply select the first $N$ samples, where $N = 300k$ or $600k$.

\textbf{Finetuning Datasets.} The details about finetuning datasets is presented in \cref{Finetuning_data}.

All pretraining and finetuning data used in the paper are publicly available.

\begin{table}[t] \centering
\caption{Size of Finetuning datasets }
\label{Finetuning_data}
\vskip 0.15in
\begin{center}
\begin{tiny}
\begin{sc}
\begin{tabular}{lccccccr}
\toprule
Benchmarks & Train & Validation & Test \\

\midrule
NLCodeSearch   & 251K & 9.6K & 19K \\
Clone Detection & 900K & 416K & 416K \\
Type Inference & 9M & 201K & 224K \\
SQA & 251K & 9.6K & 19K \\
Devign & 21K & 2.7K & 2.7K \\
DiverseVul  & 419K & 52K & 52K \\
Complexity Prediction  & 3613 & 452 & 452 \\
\bottomrule
\end{tabular}
\end{sc}
\end{tiny}
\end{center}
\vskip -0.1in
\end{table}

\section{Other Training Setup and Details}
In this section, we briefly mention different training setups we experimented with and their impact on performance.

\begin{table*}[ht]
\caption{Hyperparameters for CodeSSM, BertCoder, RoCoder and Zamba models}
\label{hyperparameter}
\begin{center}
\begin{tiny}
\begin{sc}
\begin{tabular}{lcccr}
\toprule
Hyperparameters & CodeSSM & BertCoder & RoCoder & Zamba \\
\midrule
 Number of Layers & 12 & 12 & 12 & 14 \\
 Hidden Dimension & 1024 & 1024 & 1032 & 1024 \\
 Positional Embedding  & No & Yes & Yes & No \\
 Pretraining context length & 256 & 2048 & 256 & 256 \\

\bottomrule
\end{tabular}
\end{sc}
\end{tiny}
\end{center}
\vskip -0.1in
\end{table*}

\textbf{Learning Rate (LR) scheduler.} We experimented with cosine and linear LR schedulers for both pretraining and finetuning. The CodeSSM performed the best with cosine scheduler in pretraining and linear scheduler in finetuning

\textbf{Warm-up.} Warm-up is typically used with transformers to prevent over-fitting to data seen in the early stages of training. We first experimented with no warm-up and a warm-up step of of 1\% of total training steps. We found that CodeSSM performed better on MLM validation with no warm-up. Subsequently, we experiment with a fixed warm-up step of 300 and surprisingly, with this small warm-up step, the pretraining performance improved. Further experiments are required to understand how warm-up impacts SSM training.

\textbf{Learning rate} We experimented with learning rates in proximity to what is used for transformers on code intelligence tasks. In general, SSMs performed better with lower learning rates compared to those for transformers. We used the learning rate of $5e^{-5}$ for pretraining for all the parameters (including SSM parameters). 

\textbf{Weight Decay} The weight decay for SSMs is set to 0. The weight decay for Bias and Normalization layer is also set to 0. For the rest weight decay is set to 0.01.

\textbf{Larger batch sizes.} We used gradient accumulation to increase the batch size. However, with larger batch sizes, the CodeSSM model performed worse. This could be due to less number of gradient updates with larger batch sizes and it's possible that larger batch sizes help with larger datasets.

\textbf{Tokenizer} We use the CodeT5p-220M tokenizer \cite{codet5p} from transformers library \footnote{\url{https://huggingface.co/docs/transformers/v4.17.0/en/index}}, which has a vocabulary size of 32100, for two reasons. Since it is designed specifically for code, the CodeT5 tokenizer does not significantly increase the input context length after tokenization. Moreover, it has been used for the CodeSearchNet dataset, which we use for pretraining. 

\textbf{Weight Initialization}
The layers of the CodeSSM model are initialized using the corresponding layers of the BiGS model. BiGS is a 23 layer model. For CodeSSM with 12 layers, we discard the weights of additional layers of BiGS. 
We initialize the embedding layer of  CodeSSM randomly because its vocabulary differs from that of BiGS. The task-specific heads are also initialized randomly.

\section{Pretraining evaluation} \label{pretrain_acc}
In \cref{MLM Eval}, we present the pretraining accuracy of BertCoder and multiple CodeSSM variants on a hold-out set from the CodeSearchNet dataset.

\begin{table}[ht]
\caption{MLM accuracy of different models, trained with a context window of 2048, on validation set from CSN dataset.}
\label{MLM Eval}
\begin{center}
\begin{tiny}
\begin{sc}
\begin{tabular}{lcccr}
\toprule
Model & Accuracy\\
\midrule
BertCoder  & \textbf{84.230}  \\
CodeSSM & 80.022 \\
CodeSSM-do & 79.180 \\
CodeSSM-pos & 79.640\\
CodeF & 73.324\\
\bottomrule
\end{tabular}
\end{sc}
\end{tiny}
\end{center}
\vskip -0.1in
\end{table}

\section{Theoretical basis for improvements in performance} \label{theory}
SSMs outperform Transformers due to two characteristics of their architectural design:
\begin{figure}[ht]
\vskip 0.2in
\begin{center}
\centerline{\includegraphics[width=\columnwidth, page=1, ]{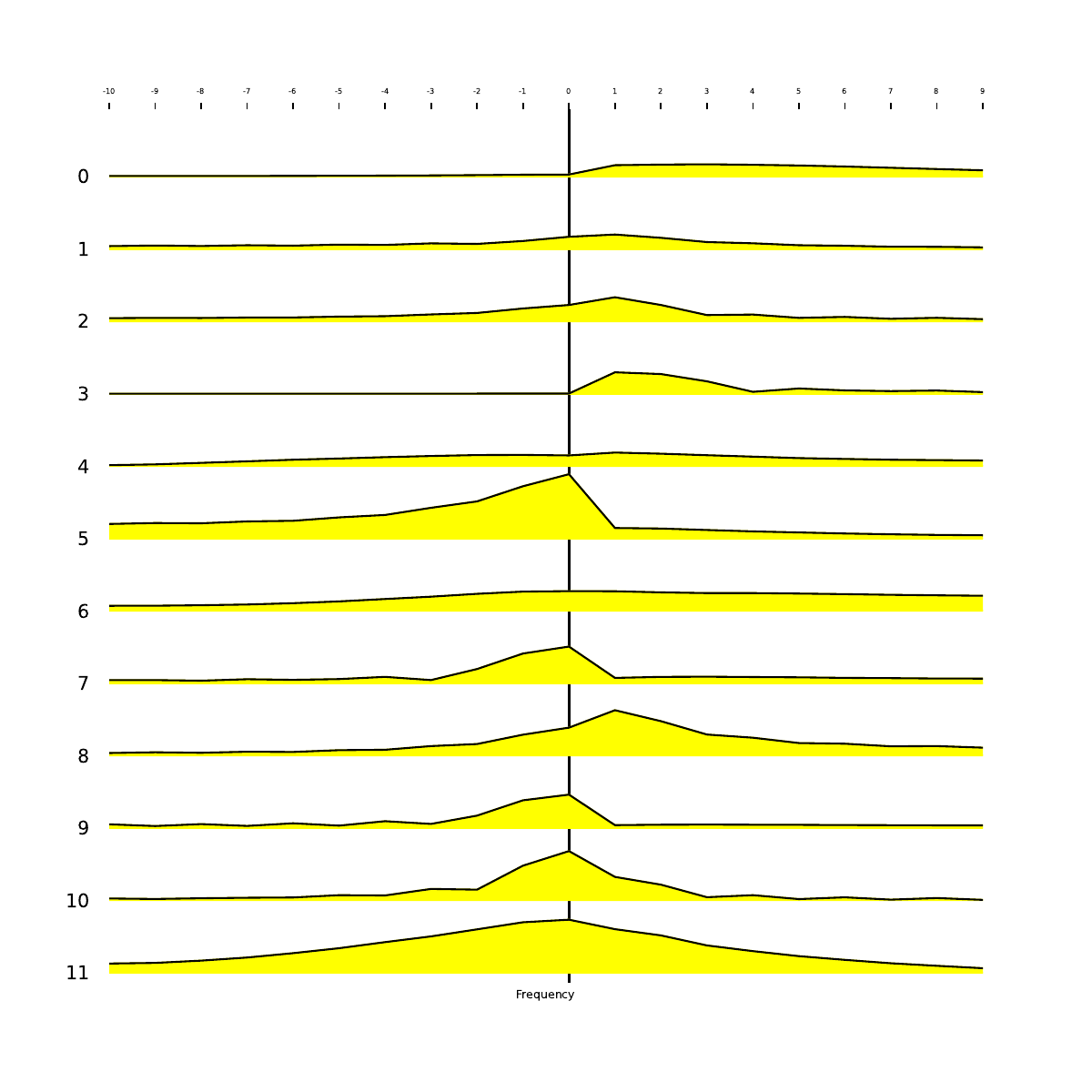}}
\caption{Magnitude Plot of Fourier Transform of SSM kernel for 12 layers of CodeSSM}
\label{mag}
\end{center}
\vskip -0.2in
\end{figure}

\begin{figure}[ht]
\vskip 0.2in
\begin{center}
\centerline{\includegraphics[width=\columnwidth, page=1, ]{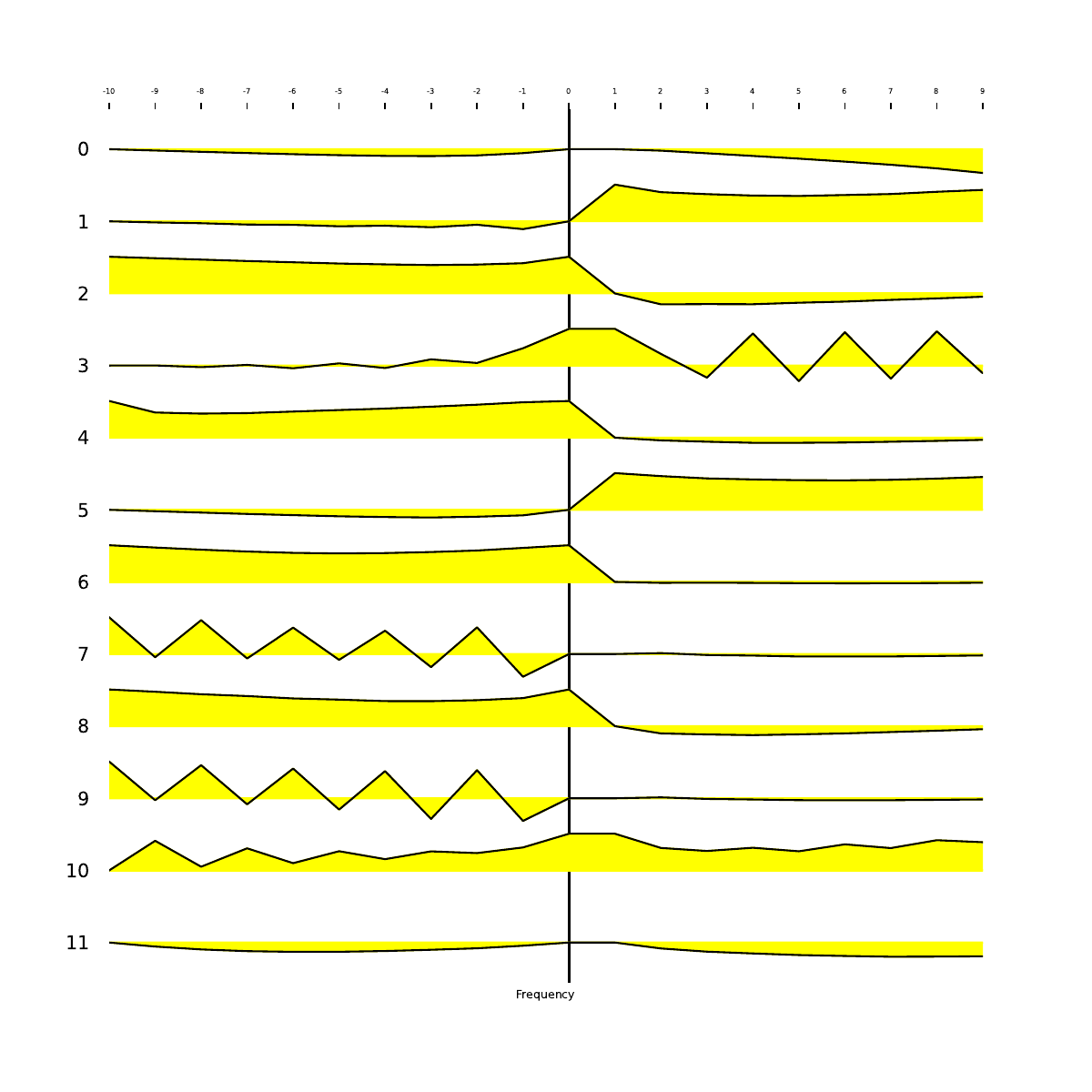}}
\caption{Phase Plot of Fourier Transform of SSM Kernel for 12 layers of CodeSSM}
\label{phase}
\end{center}
\vskip -0.2in
\end{figure}
\begin{itemize}

 \item \textbf{The initialization of A matrix} Theoretically, the initialization with HiPPO matrix allows SSMs to include unlimited historical information from a sequence. Additionally, with bi-directionality, the CodeSSM is also able to encode future information with no limit. Practically, the unlimited historical information (or context) means the input length. As we show in our paper, the pretraining length does not limit this ability and SSMs can extrapolate to longer lengths after pretraining.

 \item \textbf{Transfer function of SSMs} Given that SSMs are Linear Time-Invariant (LTI) systems, we can look at their transfer function to understand how the input affects the output. We can write this relation as:
\begin{equation}\label{ssm_tf}
    \mathbf{Y} = \mathbf{H(\omega)} * \mathbf{U(\omega)}
\end{equation}
where $\mathbf{Y}$ represents the output, $\mathbf{H(\omega)}$ and $\mathbf{U(\omega)}$ represents the Fourier transforms of the SSM kernel and the input respectively.
Since, Fourier Transforms are complex, 
\begin{equation}
\begin{split}
    \mathbf{H(\omega)} = \mathbf{M} \angle \theta \\
    \mathbf{M} \in [0, \infty] \\
    \theta \in [0, 180]
    \end{split}
\end{equation}

We can write the self-attention equation in a similar way,
\begin{equation}\label{attn_tf}
    \mathbf{Y} = \mathbf{A(u)} * \mathbf{T(u)}
\end{equation}
where $\mathbf{A(u)} = Softmax(QK^T/\sqrt{d})$ and $\mathbf{T(u)} \in [0,1]$ which is some transformation of the input. 

In SSMs, a phase shift of 180 degree between kernel and input results in forgetting some information (when added to the input through skip connection), while unbounded magnitude allows SSMs to weigh some important information very strongly. On the other hand, the transformers cannot forget any input information even if those are not relevant for a task.

\end{itemize}
The plots of magnitude and phase of all layers of the pretrained CodeSSM model for a kernel of length 10 is shown in \cref{mag} and \cref{phase} respectively. The plot shows forward kernel on the left and backward kernel on the right.

Do note, however, that $\mathbf{A(u)}$ explicitly calculates relation between each pair of input tokens, while $\mathbf{H(\omega)}$ does not. This results in a key limitation of SSMs which is that they fail to encode complex local relations. This failure can result in slight degradation in performance on tasks requiring local information with complex relations among tokens, as evidenced by performance on Type Inference (\cref{codexglue-table}). 

\begin{lstlisting}
let itemStatus={intervalId:null,
        status:"Uploading",
        urn:"",
        derivativeUrn:""};
        item.formData=itemStatus;
        };
    ...
    
let derivativeUrn=jobStatus1
       .derivatives
       .filter(derivative)
       =>derivative
       .outputType==="ifc")[0]
       .children[0].urn; 
    
\end{lstlisting}

\section{Qualitative Analysis of Failure cases in Type Inference} \label{failurecase}
We investigated the failure cases of CodeSSM on the type inference task and found that CodeSSM often fails at predicting the type when the local context is very complicated.
For example, consider the code line in the listing above.

Here the type of \texttt{derivativeUrn} is string but CodeSSM predicted it as Output while BertCoder predicted it correctly. In the same code however, CodeSSM predicted the correct type for string variables when the local context was simpler. Incorrect prediction of string type in complex contexts is also the reason for lower Top 100 F1.

To predict the correct type, the model must look at the location from where the value of the variable is coming, i.e, the model should encode which tokens influence the value of a given token. Each SSM kernel is a mixture of multiple wavelengths. The influence of one token on another depends on the frequency coefficients at a specific wavelength \cite{FoPE}. Since our SSM implementation has only one kernel per SSM block, there are only few wavelengths with high coefficients. In \cref{mag}, it can be observed that the later layers has higher coefficients for lower frequency (longer wavelengths) and hence, focus on long range dependencies. Since, type inference require both long and short range dependency, the model fails at encoding sufficient information. Including multiple kernels in SSM blocks can potentially alleviate this issue.

\section{Additional Results}
Here are some additional results on Type Inference (\cref{type inf}) with Top 100 F1 and on Clone Detection (\cref{clone}) with Precision and Recall.
\begin{table}[ht]
\caption{Results on type inference in terms of overall F1 and top-100 F1. The best performance is in bold and other noteworthy results are underlined.}
\label{type inf}
\vskip 0.15in
\begin{center}
\begin{tiny}
\begin{sc}
\begin{tabular}{lcccr}
\toprule 
Model & Context Length & Overall  F1 & Top 100 F1\\
\midrule
CodeBERT   & 512 &  \underline{0.595} & \textbf{0.850} \\
BertCoder & 2048  & \underline{0.666} & \underline{0.816}  \\
RoCoder & 256 & \textbf{0.687} &\underline{84.17} \\
CodeSSM-base  & 256 & 0.588 & 0.769\\
CodeSSM-base  & 512 & 0.603 & 0.772 \\
CodeSSM-base  & 2048 &  \underline{0.615} & \underline{0.792}\\
CodeSSM-pos & 2048 & 0.594 & 0.766 \\
CodeSSM-do & 2048 & 0.609 & 0.771 \\
CodeSSM-Sc  & 2048 &  0.619 & 0.780 \\
CodeSSM-Sc-large  & 2048 & 0.637 & 0.800 \\
\bottomrule
\end{tabular}
\end{sc}
\end{tiny}
\end{center}
\vskip -0.1in
\end{table}

\begin{table}[h]
\caption{Results on code clone detection in terms of Precision, Recall and F1 score. The best performance is in bold and other noteworthy results are underlined.}
\label{clone}
\vskip 0.15in

\begin{center}
\begin{tiny}
\begin{sc}
\begin{tabular}{lcccr}
\toprule
Model & Context Length & Precision & Recall & F1\\
\midrule
CodeBERT   & 512 &   0.947 & 0.934 & 0.941\\
BertCoder & 2048  & 0.928 & \underline{0.958} & 0.942\\
RoCoder & 256 & 0.934 & 0.944 & 0.938 \\
CodeSSM-base  & 256 & 0.888 & 0.929 & 0.908\\
CodeSSM-base  & 512 & \underline{0.949} & \underline{0.942} & \textbf{0.946}\\
CodeSSM-base  & 2048 & 0.940 & 0.926 & 0.935 \\ 
CodeSSM-pos & 2048 & 0.921 & \textbf{0.959} & 0.939 \\
CodeSSM-do & 2048 & 0.926 & 0.937 & 0.931 \\
CodeSSM-Sc  & 2048 & \textbf{0.950} & 0.926 & 0.938 \\ 
CodeSSM-Sc-large  & 2048 & 0.933 & 0.942 & \underline{0.940} \\ 

\bottomrule
\end{tabular}
\end{sc}
\end{tiny}
\end{center}
\vskip -0.1in

\end{table}

\end{document}